\documentclass[aps,prl,amsmath,amsfonts,twocolumn]{revtex4-1}%
\usepackage{mathtools}
\usepackage{verbatim}
\usepackage{color}
\usepackage{graphicx}
\usepackage{braket}
\usepackage{amsthm}
\usepackage{amssymb}
\usepackage{ulem}
\usepackage{tikz}

\newcommand{\grn}{\color{green}}
\newcommand{\red}{\color{red}}
\newcommand{\blu}{\color{blue}}
\newcommand{\blk}{\color{black}}
\newcommand{\old}{\color{black}}

\usepackage{hyperref}
\begin{document}

\title{Holographic rainbow networks for colorful Motzkin and Fredkin spin chains}

\author{Rafael N. Alexander$^{1,2}$, Amr Ahmadain$^{2}$, Zhao Zhang$^{2}$ and Israel Klich$^{2}$}
\affiliation{$^{1}
$ Center for Quantum Information and Control, University of New Mexico, Albuquerque, New Mexico 87131-0001, USA, 
\\
$^{2}$ Department of Physics, University of Virginia, Charlottesville, Virginia 22903, USA
}
\begin{abstract}
We present bulk tensor networks that exactly represent the ground states of a continuous family of one-dimensional frustration-free Hamiltonians.  These states, which are known as area-deformed Motzkin and Fredkin states, exhibit a novel quantum phase transition. By tuning a single parameter, they go from a phase obeying an area law to a highly entangled ``rainbow'' phase, where the half-chain entropy scales with the volume.
Using the representation of these ground states as superpositions of random walks, we introduce tensor networks for these ground states where local and global rules of the walker are baked into bulk tensors, thereby providing an efficient description of the ground states (some of which satisfy a volume law scaling of entanglement entropy).
\end{abstract}


\maketitle
Tensor networks can offer efficient descriptions of quantum states of interest. This is the key to their utility for constructing trial wave functions used as variational ansatz for the ground states of lattice Hamiltonians. They have been used to numerically study the behavior of correlations, entropy, and many other properties of quantum phases of matter (for a review see, e.g., \cite{Orus2014}). 
Beyond their utility for numerical studies, they offer a convenient framework for classifying the complex structure of correlations of wave functions~\cite{bridgeman2017hand} and foster connections with coarse-graining methods, such as renormalization, and related topics in field theory, such as gauge-gravity duality \cite{swingle2012entanglement}.
{\it Matrix product states are} a particularly simple class of 1D tensor networks used in the {\it density matrix renormalization group} (DMRG) procedure \cite{White1992}, successfully used in the numerical investigation of quantum phases in 1D.

Another class of tensor network states, specially tailored to describe scale-invariant systems, are represented by the {\it multi-scale entanglement renormalization ansatz} (MERA) \cite{Vidal2007ER,Vidal2008MERA}. MERA is used to represent approximate ground states of 1D quantum spin chains at criticality described by 2d conformal field theory (CFT)\cite{pfeifer2009entanglement}. 
The scale-invariance of the MERA network turned out to also play a special role in connecting it to holographic duals in the sense of the AdS/CFT correspondence~\cite{swingle2012entanglement}. Here, the bulk of a MERA tensor network can be understood as a discrete realization of 3d anti-de Sitter space ($ AdS_{3} $), identifying the extra holographic direction with the renormalization group (RG) flow in the MERA~\cite{swingle2012entanglement}.

We stress that the above treatments deal with trial wave functions, which are approximate solutions of the actual ground states. Moreover, away from 1D gapped or conformal critical points, where the MPS and MERA have been extensively studied, relatively little is known. In particular, the interpretation and use of gauge-gravity dualities beyond CFTs is not very well understood and still under intense investigation. Thus, it is of great interest to find insightful examples for tensor networks that describe exactly ground states of short-range Hamiltonians beyond MPS and MERA. 

In this Letter, we present a first example of an {\it exact} continuous family of tensor networks that describe ground states of short-range local Hamiltonians across a phase transition from area law to volume entanglement scaling.  These allow us to observe regimes associated with entanglement entropy ranging from bounded and logarithmic all the way to extensive. %
Our result is a complimentary construction to a recent example \cite{alexander2018exact} of a scale-invariant tensor network for the colorless version of the models described here, with a critical point featuring a transition between area law states through a critical point with a logarithmic entropy scaling. 
 
It is important to note that in the special case of CFTs, an explicit construction of a type of exact holography was recently achieved by mapping the Hamiltonian of free fermions on a circle onto a ``bulk" free fermion Hamiltonian inside the disc with a hyperbolic metric \cite{qi2013exact}. From the point of view of tensor networks, a related series of approximate constructions of a MERA for free fermions was shown in  Ref.~\cite{Evenbly2016, haegeman2018rigorous}. 

Our tensor network describes the ground states of the area-deformed Motzkin and Fredkin models. Motzkin models have been introduced as a new class of exactly-solvable ground states of frustration-free quantum spin chain Hamiltonians ~\cite{bravyi2012criticality,Movassagh2016Power}. A model with a similar behavior based on Fredkin gates has been introduced in Ref.~\cite{salberger2016fredkin}. 
 The Motzkin model represents an example of systems not described by a CFT and thus, present a playground where new ideas that go beyond the MERA can be explored. Moreover, they admit a class of solvable deformations, the area deformed Motzkin model, discovered in Ref.~\cite{Zhang2017novel}, with a new phase transition and simple geometric interpretation. Further studies of Motzkin and Fredkin models have explored their possible relation to non-CFT field theories~\cite{Chen2017MotzkinLifshitz}, the framework of symmetric inverse semigroups~\cite{FreeMotzkin}, and approximate quantum error-correcting codes~\cite{brandao2017quantum}. %

\begin{figure}
        \includegraphics[width=1.0\linewidth]{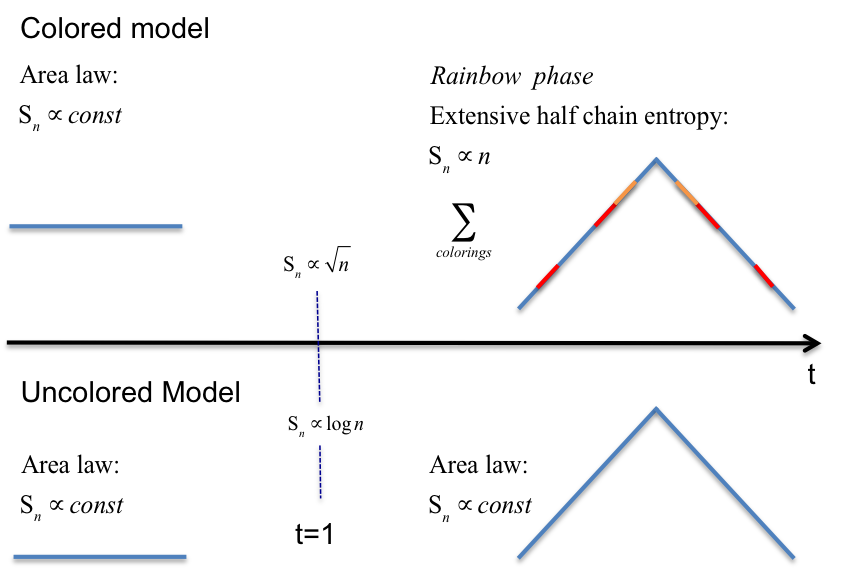}
        \caption{Phase diagram for the area deformed Motzkin Model.} %
        \label{fig:motzkinphase}
    \end{figure}

Fig.  \ref{fig:motzkinphase}  describes schematically the remarkable quantum phase diagram of this model: As a function of the parameter $ t $, the  area-deformed Motzkin model may be tuned all the way between a gapless critical phase with volume law entropy scaling, to a gapped phase obeying an area law, passing through critical points obeying logarithmic or square root entanglement scaling depending on the size of the local Hilbert space. The model at hand has a geometrically appealing description that relates 1D wave function amplitudes to objects in a 2D space that makes the establishment of the holographic tensor networks possible.

Of particular interest is the high ``half-chain" entropy phase, the ``rainbow" phase, where the state approximates a superposition of concentric entangled pairs about the middle of the system. The highly non-local nature of the rainbow phase precludes its local efficient description in terms of either MPS or MERA, and necessitated developing the new network that we present here. Apart from the deformed Motzkin and Fredkin models, which are translationally invariant in the bulk, a rainbow type ground state may also appear in spatially inhomogeneous models. Indeed, such a phase was first demonstrated by Vitagliano et al. in \cite{vitagliano2010volume} for a spin chain with an explicitly broken bulk transitional invariance via exponentially varying coupling constants. The concentric singlet phase was shown in the strong coupling limit, \cite{vitagliano2010volume} where the model was analyzed using a Dasgupta$-$Ma real-space renormalization group technique, and also studied in \cite{ramirez2014conformal} and via mapping to free fermions and exact diagonalization.

{\it Motzkin walks and Motzkin ground states.} The Motzkin model is a one-dimensional spin-$j$ chain ($j$ integer). For ${j=1}$, identifying each local spin basis state $\{ \ket{1}, \ket{0}, \ket{-1}\}$ with a line segment $\{ \diagup,  \text{\textbf{---}}, \diagdown \}$, respectively, allows us to represent states as a superposition of walks. 
Higher dimensional spins can be analogously defined using colored walks. For example,  for $j=3$,  the basis states $\{ \ket{3}, \ket{2}, \ket{1}, \ket{0}, \ket{-1}, \ket{-2}, \ket{-3}\}$ are identified with $\{ \grn \diagup \blk, \blu \diagup\blk, \red\diagup\blk,  \text{\textbf{---}}, \red \diagdown \blk, \blu \diagdown \blk, \grn \diagdown \blk\}$, respectively. 

The Motzkin model has a unique, zero-energy frustration free ground state. For the spin $j>1$ ($j=1$) case, this ground state is a superposition of walks called ``colored (uncolored) Motzkin walks''. A Motzkin walk $w$ is a walk on the $\mathbb{Z}^{2}$ lattice using the line segments $\{ \diagup,  \text{\textbf{---}}, \diagdown \}$ that starts at $(0,0)$, goes to $(2n,0)$, and never goes below the $y=0$ line. In the colored walks, all upwards steps have an arbitrary color. However, the color of a downward step $(k,m)\rightarrow(k+1,m-1)\in w$ must match the color of the most recent upwards step occurring at the same height; i.e., $color((k,m)\rightarrow(k+1,m-1))=color((l,m-1)\rightarrow(l+1,m))$ where $l=\text{max}(l^{\prime})$ s.t. $l^\prime <k$ and $(l^{\prime},m-1)\rightarrow(l^{\prime}+1,m) \in w$. Denoting the set of colored Motzkin walks with $c$ colors on $2n$ steps $\mathcal{M}^{2n}_c$, the ground state can be written as:
 \begin{equation} |\Psi(t)\rangle = \frac{1}{\mathcal{N}}\sum_{\substack{w \in \mathcal{M}^{2n}_c}} t^{\mathcal{A}(w)} |w\rangle. \label{cgs} \end{equation} 
 Here $\mathcal{A}(w)$ denotes the area below the Motzkin walk $w$, and $\mathcal{N}$ is a normalization factor.
A similar type of ground state occurs in the Fredkin models, which are half-integer spin models that have essentially the same structure, but without the ``flat'' move.
More details can be found in references \cite{salberger2016fredkin,salberger2017deformed,zhang2017entropy, udagawa2017finite}.
The half-chain entropy is easily understood from observing the dominant Motzkin walks in the limits of $t\rightarrow \infty$ and $t\rightarrow 0$ and is described in Fig. \ref{fig:motzkinphase}. The deformed Motzkin and Fredkin walks can naturally be viewed as constrained trajectories of a random walker in the presence of drift, with the ``x'' axis playing the role of time.

 \begin{figure*}
\includegraphics[width=1\linewidth]{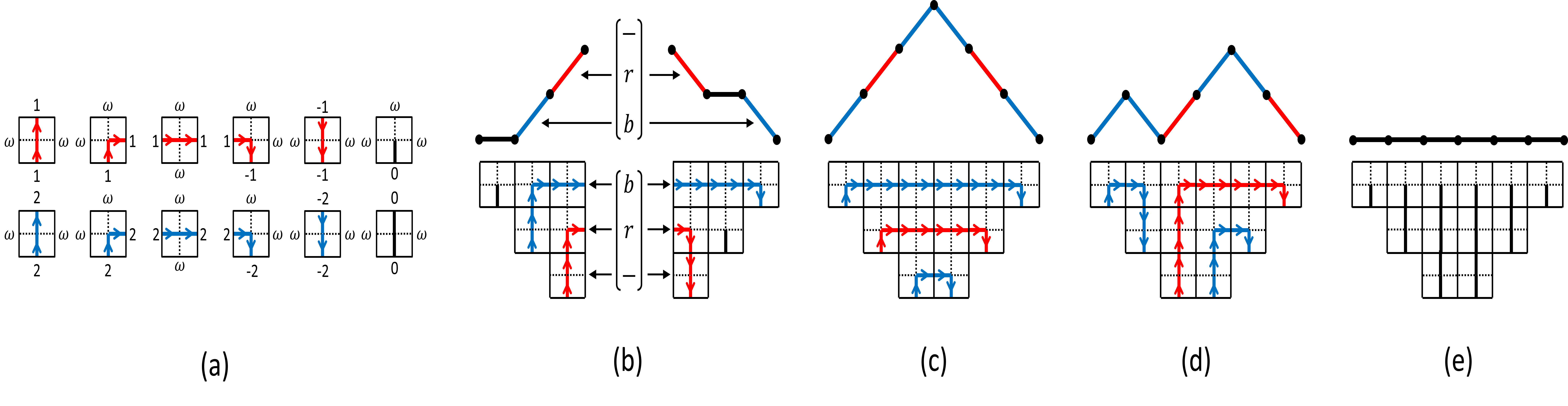}
\caption{$\mathbf{(a)}$ 12 tiles for the spin 2 case. In the spin-$j$ case, there will be $j$ differently colored copies of each arrowed tile (there are five distinct arrowed configurations). 
$\mathbf{(b-e)}$ Examples of the mapping between colored walks (top) and full packing of paths in the bulk of the network (bottom). The sequence of paired line segments along a vertical cut in the walk (top) mirrors the pairing of walk segments across the same in the grid (bottom). This is shown explicitly in (a).}\label{fig:TNtiles}
\end{figure*}

{\it The rainbow tensor network representation of the area-weighted colored Motzkin ground state}.\label{GeneralExactTN} 
As discussed in the previous section, the ground state of the spin-$j$ Motzkin model is related to the time-evolved path of a random walker with $2j+1$ choices at every time step, along with a global constraint that the path's height must never be negative. Motivated by this correspondence, we introduce a tensor network representation for such ground states where these rules are baked right into the building blocks. 
We find that the network can be graphically represented as a collection of possible tilings showing colored arcs or ``rainbows".

{\it Walks as tiles}. Consider the set of tiles shown in Fig.~\ref{fig:TNtiles}~(a). These can be used to tile a square lattice. We say that a tiling is {\it valid} if the edges of each tile match, and if the following boundary conditions are satisfied: all upwards, left, and right facing boundary edges must take the value $\omega$, and all downwards facing boundary edges are prohibited from taking the value $\omega$. 

\begin{figure}
\includegraphics[width=1.\linewidth]{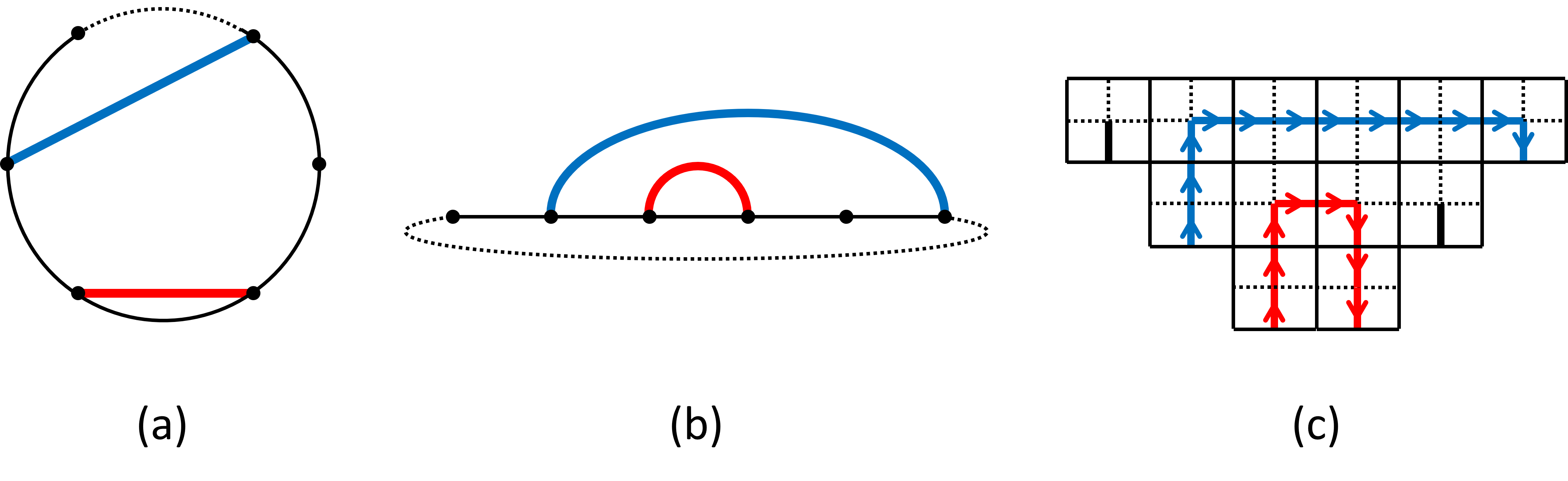}
\caption{Length $2n$ Motzkin walks are one-to-one with non-intersecting chords between $2n$ points on a circle. These, in turn, are one-to-one with full packings of the interior of an inverted step pyramid by paths of arrowed and arrowless lines. } \label{fig:arctn} %
\end{figure}

The set of length $2n$-colored Motzkin walks is isomorphic to the set of valid tilings of an inverted step-pyramid. Examples are shown in Fig.~\ref{fig:TNtiles}~(b-e). Each valid tiling corresponds to a full-packing of the interior of the square grid by non-intersecting arrowed and arrowless paths. The arrowed paths begin traveling straight upwards from the bottom of the grid, take two right turns (following a $\Pi$-shaped path), then return to another location at the bottom of the grid. The arrowless lines form straight vertical paths from the bottom of the grid and terminate in the interior. Each colored length-$2n$ Motzkin walk is isomorphic to a configuration of colored non-intersecting chords that join $2n$ points that lie on a circle. See Fig.~\ref{fig:arctn}(a) for an example. Flattening the circle---as shown in Fig.~\ref{fig:arctn}(b)---results in a configuration of nested colored arcs. These are ``smoothed'' counterparts of configurations of $\Pi$-shaped paths that pack the square grid, as shown in Fig.~\ref{fig:arctn}(c). Each tiling is uniquely specified by the numerical values on the bottom horizontal edge of each column (recall that the value $\omega$ is prohibited). For the tiling to be valid, each path seeks to maximize the height it reaches in the interior. 

{\it Tiles as tensors}. We introduce a tensor network that is designed to sum over all valid tilings. Thus, it represents a sum over all Motzkin walks, and hence, the ground state of the Motzkin model. 

\begin{figure}
\includegraphics[width=1 \linewidth]{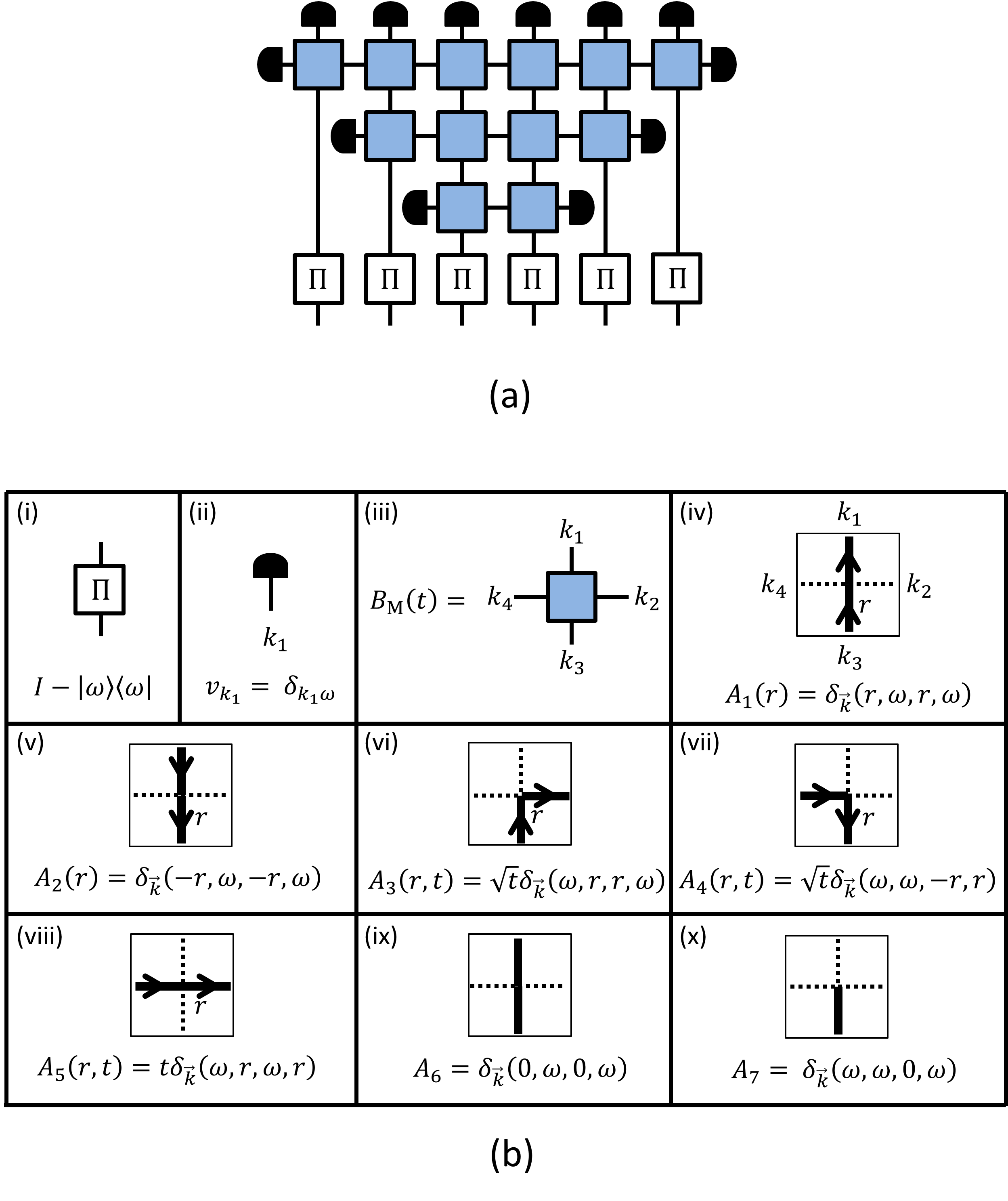}\caption{$\mathbf{(a)}$ In this tensor network the $2n$ physical spins are located on the lower boundary. $\mathbf{(b)}$ \textbf{(i)} Projector onto the spin-$j$ subspace. \textbf{(ii)} Unit vector in the (non-physical) $\omega$ direction. \textbf{(iii)} The four-index tensor $B$ (defined in Eq.~\ref{eq:BMdef}). \textbf{(iv-x)} Rather than showing colors explicitly (as in Fig.~\ref{fig:TNtiles}(a)), we have labeled the tiles containing arrowed lines with the variable $c\in \{1, \dots j \}$. The $A$ tensors define a one-to-one mapping from tile edges to contractions with $\ket{-c}, \ket{0}, \ket{c},$ or $\ket{\omega}$.}\label{fig:invtriangle} %
\end{figure}

The tensor network is shown in Fig.~\ref{fig:invtriangle}(a). It is a two-dimensional square lattice embedded within an upside-down step pyramid. Physical indices are arranged along the bottom edge. The basic building block is the four-index tensor $B$, defined as 
\begin{align}
B(t) \coloneqq \sum_{l=1}^{5} \sum_{c=1}^{j}  A_{l}(c, t) + A_{6} + A_{7} \label{eq:BMdef}
\end{align}
where the $A_i$ are rank-1 tensors with four indices, as shown in Fig.~\ref{fig:invtriangle}(b). These are defined via Kronecker delta $\delta_{jk}$,
\begin{align}
\delta_{\vec{j}}(w, x, y, z)\coloneqq \delta_{j_{1} w}\delta_{j_{2} x}\delta_{j_{3} y}\delta_{j_{4} z}. \label{eq:delta}
\end{align}
Contracting a single index between two $B$ tensors corresponds to summing over tile configurations that match on the joining edge. 

To ensure that the boundary conditions of a valid tiling are met, we contract the left, right, and upwards facing boundary legs with the vector $\ket{\omega}$, and the downwards facing boundary legs with the projector $\Pi= I - \ket{\omega}\!\!\bra{\omega}$ (see Fig.~\ref{fig:invtriangle}(a)). 

{\it Encoding of correlations in the tensor network.} The horizontal virtual bonds across a cut between two columns of $B$ tensors store an ordered list of colors corresponding to unpaired walk segments across that cut in the walk.  To see this, consider a cutting a given walk between sites $z$ and $z+1$ at a height $h_{z}$. Then, exactly $h_{z}$ pairs of locations split by this cut are perfectly correlated in the color degree of freedom. Denote these pairs of sites by $(x_{1}, y_{1}) \dots (x_{h_{z}}, y_{h_z})$, where $x_j\leq z<y_j$, $\forall j$. Assume these are ordered so that $x_{1} < x_{2} < \dots x_{h_z}$ (and therefore, $y_{h_z} < y_{h_z-1} < \dots y_{1}$), and denote the colors of these pairs by $c_{1}, c_{2}, \dots c_{h_z}$. Correlations across the cut are encoded within these color configurations. This type of data structure is known as a {\it stack}; the upwards steps to the left of the cut are ``undone" by a downwards steps to the right of the cut in reverse order. In the tensor network, the colors are stored in order ($c_1$, $c_2$, $\dots$) from top to bottom moving downwards along the cut. See Fig.~\ref{fig:TNtiles}(b) for an example. 
 
{\it Dependence on t.} The $B$ tensor in Eq.~\ref{eq:BMdef} explicitly includes the $t$ parameter. The tiles $\{A_{l} \}$ have a factor of $\sqrt{t}$ for every horizontal arrow segment that appears. As discussed above, the height/color information of each walk is stored in the horizontal virtual bonds between two columns of $B$ tensors. Therefore, for a given tiling, each horizontal arrow segment contributes half a unit of area. Scaling the tiles by the number of horizontal arrows, therefore, corresponds to scaling the walks by the number of area units they cover. 

{\it Fredkin ground state.} Recall that the ground states of the Fredkin model can be represented by walks that do not include any horizontal segments. The tensor network shown in Fig.~\ref{fig:invtriangle} (a) can be re-purposed for such models if all tiles containing solid lines without arrows are removed from $B$. The new tensor is defined as 
\begin{align}
B^{\prime}(t) \coloneqq \sum_{r=\tfrac{1}{2}}^{j} \sum_{l=1}^{5} A_{l}(r, t).
\end{align}
In addition, we must map nonzero integer spin values $j$ to half-integer values $j/2$. The correspondence between walks and arrowed paths in the network is otherwise identical to the Motzkin case. Examples showing the correspondence between walks and tiles are (c) and (d) in Fig.~\ref{fig:TNtiles}.

The rainbow tensor network is an exact representation for any member of the family of colored and area-weighted Motzkin or Fredkin ground states. It provides an efficient description of any such state (only $O(n^2)$ many identical four index tensors are required to specify it). However, if each column of square tensors is contracted and the horizontal legs are appropriately ``fused", then it yields an MPS with bond dimension that grows exponentially with the system size.  %

 Beyond tensor networks for 1D quantum spin chains, such a geometric approach has been useful in many problems of statistical mechanics including quasi-crystals spin glasses \cite{Garrahan2009Molecular}, dimer models and spin jams \cite{klich2014glassiness}  where the building blocks, the tiles, correspond to allowed local physical configurations and offers a convenient, graphical approach to tensor networks where a geometrical picture of the state is involved.
 
 {\it Holography and tensor networks.} The construction of a homogeneous MERA tensor network is special in ways that do not always extend to systems without scale invariance or logarithmic scaling of entanglement entropy. In a MERA, tensor elements obtained numerically are {\it generic}, therefore, generic correlations between a pair of operators acting at positions $x_{1} $ and $ x_{2} $ are carried through the bonds/links of the network \cite{Evenbly2011TNGeometry},  giving:
    \begin{equation} \label{MERACorrelation}
 G(x_{1},x_{2}) \approx  e^{-\alpha D(x_{1},x_{2})}
    \end{equation}
for some correlation function $ G(x_{1},x_{2}) $, where $ D(x_{1},x_{2}) $ is the graph distance (i.e. minimal number of edges) between $ x_{1} $ and $ x_{2} $ within the tensor network, and $\alpha$ is a positive constant that depends on the operators in question. 
Since in a MERA, $D(x_{1},x_{2}) \approx log (|x_{1} - x_{2}|)$, it follows from Eq. \ref{MERACorrelation} that $ D(x_{1},x_{2})$ dictates a power law scaling of $ G(x_{1},x_{2}) $ as expected for a CFT. From the point of view of the network structure, MERA may be naturally seen as a type of holographic description. In particular, it clearly demonstrates features such as consistency with the Ryu-Takayanagi (RT) formula \cite{ryu2006holographic}, relating entanglement entropy of a region with its minimal bounding surface in the holographic direction. For non-conformal field theories, a holographic gravity dual may not, in general, be able to simultaneously satisfy an RT-like formula for entanglement entropy and a semi-classical description of correlation functions given in terms of geodesics.
Note that, as remarked before, gap scaling show that even the colorless Motzkin system, where entropy behaves logarithmically, is not a CFT \cite{bravyi2012criticality,Movassagh2017}. Indeed, a corresponding field theory has yet to be properly described  Chen et al. used large-scale density matrix renormalization group to investigate the spectrum of low-lying excitations in a generalized Motzkin spin chain \cite{Chen2017MotzkinLifshitz}. The authors were able to find two gapless modes with different dynamical scaling exponents, $z=2.7$ and $z=3.16$ respectively, which they argued is evidence for multiple dynamics. They also constructed a continuum limit of the colorless Motzkin ground state as a ground state of a $z=2$ Lifshitz scalar field theory with orbifold boundary (Note that due to the mismatch in dynamical scaling, this $z=2$ Lifshitz field theory is insufficient to fully describe the spectrum of the Motzkin spin-chain).

What about the tensor network we presented here? Viewed as a graph, it is defined on a square grid, that seems to correspond to a ``flat" holographic metric. However, if we compute correlation functions, they are strongly dependent on position. In particular, in the $t\rightarrow \infty$ limit, correlation functions represent concentric pairs of maximally entangled pairs. The corresponding holographic geometry is perhaps more appropriately represented as an array of concentric ``wormholes" \cite{maldacena2013cool,jensen2013holographic}, i.e., a rainbow.
    
As a concrete example, consider the correlations between color degrees of freedom. We concentrate on the $t\rightarrow \infty$ limit, where the Motzkin walk is characterized by a tall triangular mountain with small corrections. In this case the ``up-down'' degree of freedom of the spins is almost frozen, however, colors are widely fluctuating via completely correlated pairs symmetric about the middle of the chain. To quantify the correlations, we will assume two colors, say red and blue. Here the local Hilbert space is $5$ dimensional, consisting of the local states ${\color{red} \downarrow, \uparrow},{\color{blue} \downarrow, \uparrow},{\color{black} -}$. 
We can define a color operator $C$, by its action: 
    \begin{align}
    C|s\rangle=color(s)|s\rangle
    \end{align}
    where $color({\color{red} \downarrow, \uparrow})=-1$, $color({\color{blue} \downarrow, \uparrow})=1$, and $color({\color{black} -})=0$ for the horizontal step. 
     In the ground state $|\Psi(t)\rangle$, consider the color-color correlation function
    \begin{eqnarray}
    G_{x_1,x_2}\equiv \langle C_{x_1} C_{x_2}\rangle-\langle C_{x_1}\rangle \langle C_{x_2}\rangle=\langle C_{x_1} C_{x_2}\rangle.
    \end{eqnarray}
Note that since the state has no particular color preference
$\langle \Psi(t) |C_x|\Psi(t)\rangle=0$ for any point $x$ and any value of $t$. If a holographic metric allows for a semi-classical description of the state we should expect:
    \begin{eqnarray}
    G_{x_1,x_2}\sim e^{-h D(x_1,x_2)} 
    \end{eqnarray}
where $D$ is the geodesic distance between $x_1$ and $x_2$ when going through the holographic geometry and $h$ is a parameter related to the scaling of our "color" operator. 
    
\begin{figure}
        \includegraphics[width=0.6\linewidth]{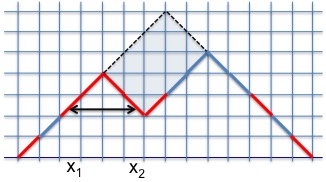} \caption{This path contributes to the color correlation of spins $x_1$ and $x_2$ with maximal area.} \label{fig:corrpath}
    \end{figure}

\begin{figure}
    \includegraphics[width=0.8\linewidth]{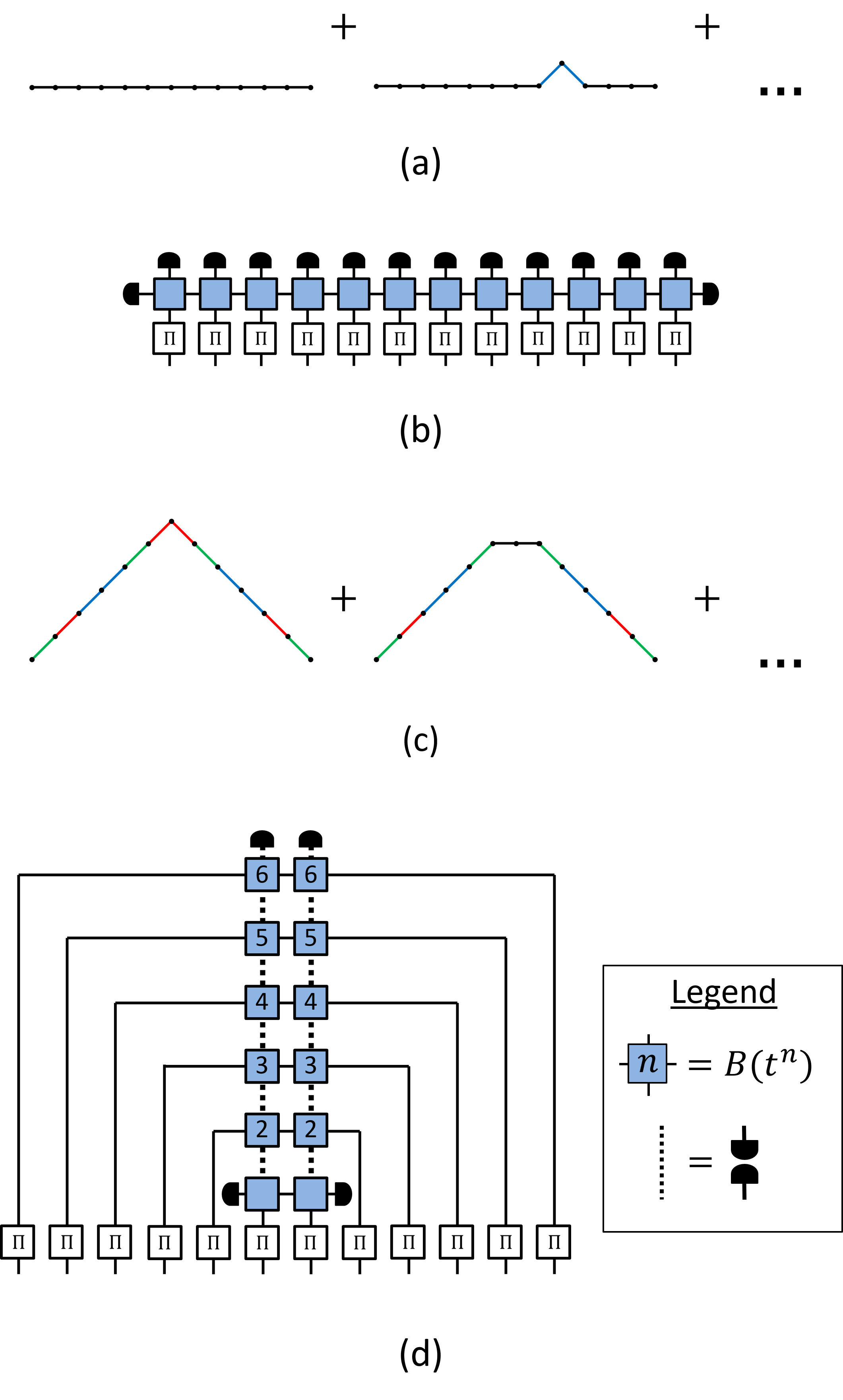}
    \caption{\textbf{(a)} For fixed $n$, in the limit of $t\rightarrow 0$, the ground state can be approximated by discarding walks that have area $\geq 2$, i.e.,  keeping only the flat walk plus all colorings of area 1 walks. \textbf{(b)} Such walks are contained within the tensor network shown, which can be viewed as a horizontally truncated rainbow tensor network. The geometry of this network is the same as an MPS. \textbf{(c)} For fixed $n$, in the limit of $t\rightarrow \infty$, the ground state can be approximated by discarding walks that have area $\leq (n^{2} -2)$, i.e., keeping only all possible colorings of the maximum height walk and area $(n^{2} -1)$ walks. \textbf{(d)} Such walks are contained within the tensor network shown, which is similar to a vertically truncated rainbow network, but where the $t$-dependence of the $B$ tensors depends on which row it appears in, and the majority of the vertical contracted indices have been replaced with projectors onto the $\ket{\omega}$ state. }
    \label{fig:truncated}
\end{figure}

How will $D(x_1,x_2)$ behave in our state? First, write:
    \begin{eqnarray}
    G_{x_1,x_2}=\frac{\sum_{w\in {\cal C}_{+}}t^{2A(w)}-\sum_{w\in {\cal C}_{-}}t^{2A(w)}}{\sum_{w}t^{2A(w)}},
    \end{eqnarray}
where we defined the sets:
    \begin{eqnarray}
    {\cal C}_{\pm}(x_1,x_2):=\{ w\in \mathcal{M}^{2n}_s : color(x_1)color(x_2)=\pm 1 \}
    \end{eqnarray}
Consider a particular Motzkin walk in ${\cal C}_{+}(x_1,x_2)$. If in this walk the color in $x_2$ is independent of that of $x_1$ (for example if in $x_2$ the step is upwards) then there will be a corresponding Motzkin walk in ${\cal C}_{-}(x_1,x_2)$ with the same area, and no contribution to $G_{x_1,x_2}$. Thus, walks where the colors of $x_1,x_2$ are not correlated will not contribute to the sum. We therefore write $G$ as:    
\begin{eqnarray}
    G_{x_1,x_2}=\frac{\sum_{w:\,\text{s.t. $color(x_1)=color(x_2)$}}t^{2A(w)}}{\sum_{w}t^{2A(w)}}
    \end{eqnarray}
In the ${t\rightarrow\infty}$ limit, the asymptotically leading contributions to the area weighted Motzkin walks are determined by maximal area walks that contribute to the correlation.
Assuming $x_2-x_1$ is even, a maximal area walk with mandatory $color(x_2)=color(x_1)$ is illustrated in Fig. \ref{fig:corrpath}. If we choose our coordinate system such that $x=0$ corresponds to the middle of the spin chain, then we get
        \begin{eqnarray}
    G_{x_1,x_2}\sim \frac{2^n t^{2A_{max}(w)}}{2^n t^{4n}}\sim e^{-h |x_1^2-x_2^2|},
    \end{eqnarray}
with $h = \log t$. In the continuum limit we will have 
\begin{eqnarray}
    G_{x_1,x_2}\propto (\delta(x_1-x_2)+\delta(x_1+x_2)),
    \end{eqnarray}
Thus, we assume that in a metric describing this state, points that are symmetrical around the middle should be connected by short geodesics. 
On the other hand, in the limit $t\rightarrow 0$, the situation is reversed - there are only correlations between very close points. Appropriate approximate tensor network are described in Fig. \ref{fig:truncated}.

\old

One interesting question in this regard is, what is the nature of a {\it bulk} Hamiltonians generating such superpositions? It is quite clear that the rainbow tensor network, for example, represents a bulk state that is unlikely to be generated by an exact local bulk Hamiltonian: it consists of a superposition of bulk rainbows that cannot be deformed to each other by local bulk moves. Is this a feature of the particular representation we found, or is it a general expectation that high entanglement ground states have to be associated with non-local bulk Hamiltonians? 

\emph{Acknowledgments.}
We thank Glen Evenbly, Ramis Movassagh, Vladimir Korepin and Xiao-Liang Qi for discussions.
The work of I.K., A.A., and Z.Z. was supported in part by the NSF grant DMR-1508245.
 The work of R.N.A. was partially supported by National Science Foundation Grant No.~PHY-1630114. 
\bibliography{MotzkinTNBib}

%%%%%%%%%%%%%%%%%%%%%%%%%%%%%%%%%%%%%%%%%%%%%%%%%%%%%

%%%%%%%%%%%%%%%%%%%%%%%%%%%%%%%%%%%%%%%%%%%%%%%%%%%%%

\appendix
\section{Motzkin Hamiltonian}
The area deformed Motzkin Hamiltonian, defined on a spin-chain with $2n$ sites reads:
 \begin{equation} H= \Pi_{boundary} + \sum_{j=1}^{2n-1}\Pi_{j} +\sum_{j=1}^{2n-1} \Pi_{j}^{cross}, \label{cHam} 
 \end{equation} 
 where  $\Pi_{j},\Pi^{cross}_{j} $ act on the pair of spins $j,j+1$ with
 \begin{align} & \Pi_{j} = \sum_{k=1}^{s}(|\Phi^k_{t}\rangle\langle\Phi^k_{t}|_{j,j+1} + |\Psi^k_{t}\rangle\langle\Psi^k_{t}|_{j,j+1} + |\Theta^k_{t}\rangle\langle\Theta^k_{t}|_{j,j+1}),\\  & \Pi_{j}^{cross} = \sum_{k \neq k'}|u^kd^{k'}\rangle\langle u^kd^{k'}|, \\  & \Pi_{boundary} = \sum_{k=1}^{s}(|d^k\rangle\langle d^k|_1 + |u^k\rangle\langle u^k|_{2n}).
 \end{align} where  $\Phi^k,\Psi^k,\Theta^k$ are the following states on pairs of neighboring spins 
 \begin{align} |\Phi^k_{t}\rangle & \propto |u^k0\rangle - t|0u^k\rangle~;~|\Psi^k_{t}\rangle \propto |0d^k\rangle - t|d^k0\rangle~; \\ |\Theta^k_{t}\rangle & \propto |u^kd^k\rangle - t|00\rangle \label{statess}\end{align} %

\end{document}